\documentclass[twocolumn,aps,prl,showpacs]{revtex4}

\usepackage{graphicx}
\usepackage{dcolumn}
\usepackage{bm}

\begin{document}

\title{Silicon-based molecular electronics}
\author{T. Rakshit, G-C. Liang, A. W. Ghosh and S. Datta}
\email[Author list is reverse alphabetical. All authors
contributed equally. Email: ]{rakshit/liangg/ ghosha/datta@ecn.purdue.edu}
\affiliation{School of Electrical and Computer Engineering, Purdue
University, W. Lafayette, IN 47907}%

\date{\today}
\widetext
\begin{abstract}
Molecular electronics on silicon has distinct advantages over its
metallic counterpart. We describe a theoretical formalism for
transport through semiconductor-molecule heterostructures, combining
a semi-empirical treatment of the bulk silicon bandstructure with a
first-principles description of the molecular chemistry and its
bonding with silicon.  Using this method, we demonstrate that the
presence of a semiconducting band-edge can lead to a novel molecular
resonant tunneling diode (RTD) that shows negative differential
resistance (NDR) when the molecular levels are driven by an STM
potential into the semiconducting band-gap. The peaks appear for
positive bias on a p-doped and negative for an
n-doped substrate.  Charging in these devices is compromised by the RTD
action, allowing possible identification of several molecular highest occupied
(HOMO) and lowest unoccupied (LUMO) levels. Recent experiments by
Hersam
{\it{et al.}} \cite{rHersamexpt} support our theoretical predictions.
\end{abstract} 
\bigskip


\maketitle

Traditionally molecular electronic efforts, both
theoretical \cite{rtheory,rDamle} and experimental
\cite{rReed,rexpt,rWeber,rLindsay,rReif}, have been driven by
thiol-gold chemistry to molecules bonded to gold substrates. However
several recent experiments have demonstrated the
feasibility of attaching various molecules on clean silicon surfaces
\cite{rsiliconexpt}. The development of molecular electronics on silicon
is particularly important for two reasons. Firstly, it will enable
the development of integrated devices that can utilize the powerful
infrastructure provided by the silicon-based IC industry. Secondly,
unlike gold, silicon has a bandgap that one can take advantage of
to design a new class of resonant tunneling devices
with possible applications in logic \cite{rFETlogic} and
low-power memory \cite{rmem}. In view of these significant potential
payoffs we believe it is worthwhile at this time to develop models
that can be used to analyze the electrical characteristics of such
silicon-based molecular devices.

The purpose of this paper is to present (i) a general formulation
suitable for modeling silicon-based molecular devices; (ii) realistic
principles for designing RTDs based on such structures; and (iii) showing
ways to map out the molecular energy spectrum that is not realizable in
normal experiments with gold contacts. Molecular devices on
silicon require a formulation that can account for the bandgap, surface
band-bending and surface reconstruction. One problem commonly encountered
is that standard quantum chemical basis sets describe molecular energy
levels well but provide a very poor description of the semiconductor
bandstructure. We present a scheme that can be used to integrate the two
distinct systems seamlessly and present results that combine ab initio
treatments of the molecule with a semi-empirical description of the
silicon bands. The same approach can be used to integrate the molecule
with more advanced treatments of the silicon substrate.

\begin{figure}[!t]
\vspace{1.4in}
{\includegraphics{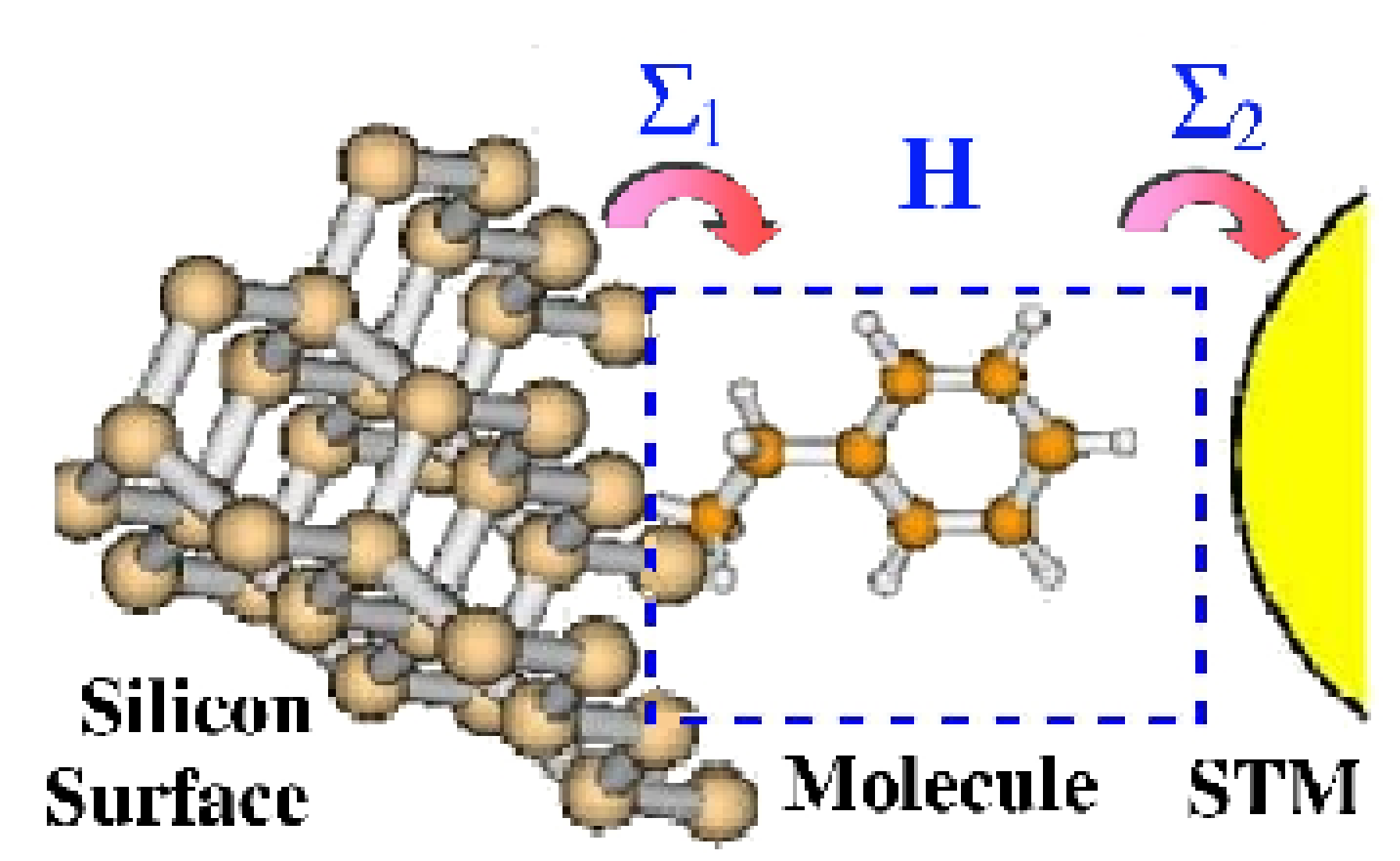}}
\vskip 1.4in
{\includegraphics{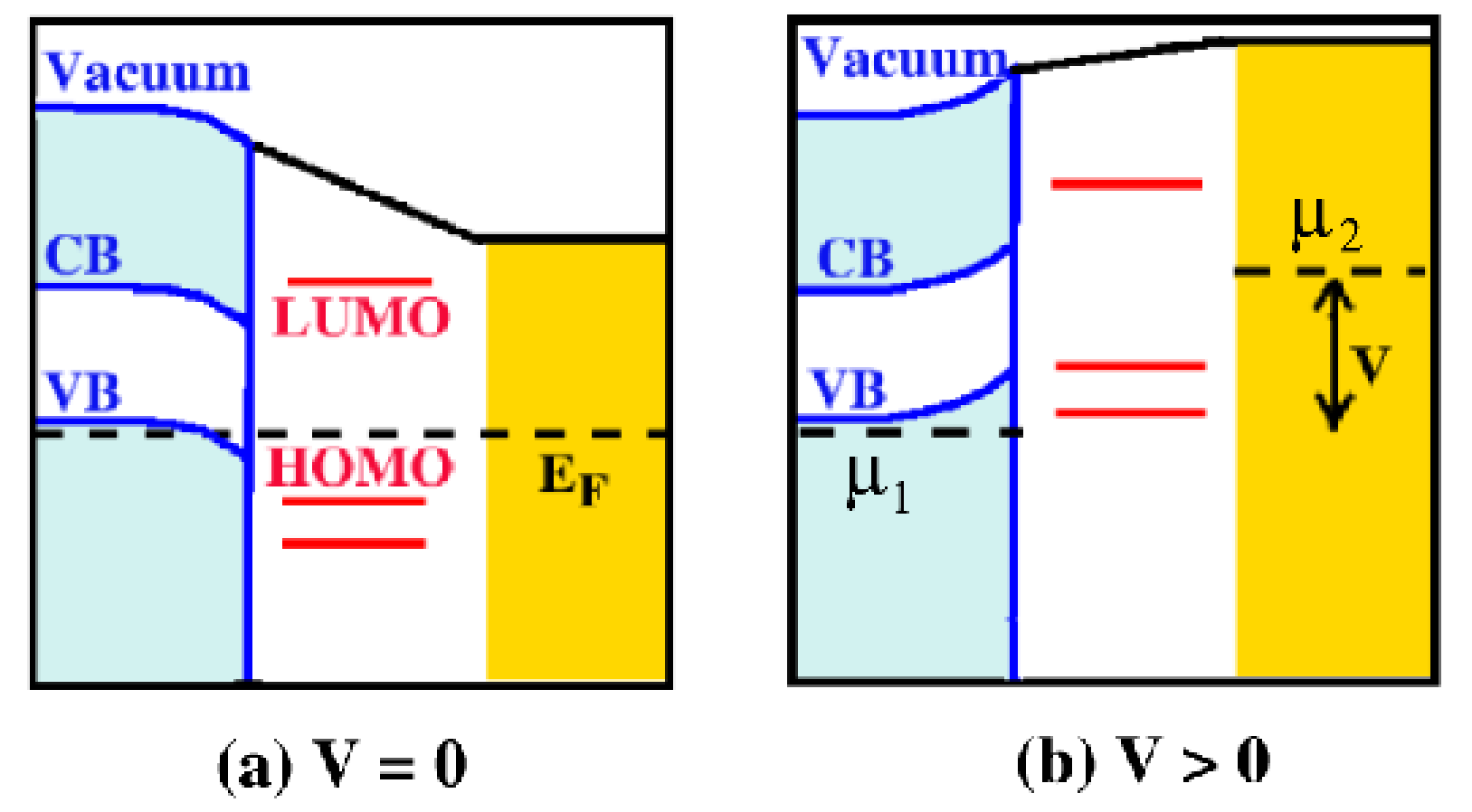}}
\caption{Schematic description of molecular RTD involving a silicon 
band-edge and an STM tip.
The extended device Hamiltonian  $H$
includes the molecule and a part of the silicon surface relevant to
the bonding with the molecule. The influence of bulk silicon and STM
contacts are incorporated through self-energy matrices $\Sigma_{1,2}$. 
Bottom: (a) Equilibrium band-alignment in  p-silicon-molecule-metal 
heterostructure corresponds to a flat Fermi energy $E_F$ (dashed line) near
the silicon valence band-edge (VB). Shown also are the
silicon conduction band (CB), the metal band (right), and the molecular
energy levels. (b) For positive substrate bias, the levels move up until the
HOMO levels leave the bulk silicon valence band into the band-gap, leading
to a sudden drop in conductance and a corresponding NDR in the I-V.}
\label{f0}
\end{figure}

Our results show that we can expect to see NDR using molecules like
styrene or TEMPO (2,2',6,6'-tetramethyl-1-piperidynyloxy) on p-silicon at
realistic positive substrate voltages, when current flow occurs through
the HOMO levels. A schematic description is shown in Fig.~\ref{f0}. Recent
experimental observations \cite{rHersamexpt} are in good agreement with
these results. By contrast, it appears that much larger negative voltages
will be needed to observe these effects on n-type silicon where conduction
takes place via the LUMO levels.  This is an important observation that
deserves careful attention, since little work has been reported to date
for electronic conduction through the LUMO levels.

\begin{figure}[ht]
\vspace{1.9in}
{\includegraphics{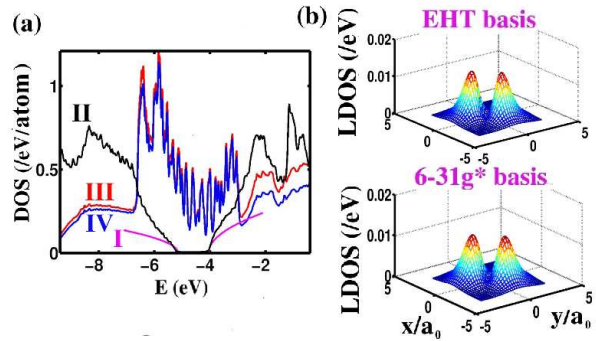}}
\caption{Results for Si(100) density of states (DOS) in a hybrid basis:
(a) bulk DOS in (I) effective mass theory and
(II) extended Huckel theory (EHT) with parameters fitted to match the bulk
silicon bandstructure \cite{rCerda}. The surface DOS match in energy domain
for (III) EHT and (IV) 6-31g$^*$ bases.  (b) The surface DOS match in
real space as well, plotted here at the Fermi energy, $a_0$ denoting the Bohr
radius.}
 \label{f0b}
\end{figure}

{\it{Formalism}}.
We will use a method by which a bulk silicon bandstructure calculation
(e.g. effective mass \cite{rlundu} or tight-binding sp$^3$s$^*$
\cite{rsp3s}) can be connected up with a detailed ab-initio treatment
of the molecule, without introducing spurious basis-transformation
related effects at their interface. We accomplish this connection
through the real-space Green's function obtained on a surface silicon
atom, which can be looked upon {\it{both as a scooped-out part of the
silicon crystal, as well as an extended part of the molecular device}}.
The connection is established as:
\begin{equation}
g(E) = \left[ES_0 - H_0 - \Sigma_1(E)\right]^{-1}
\label{e00}
\end{equation}
where $S_0$ and $H_0$ are the ab-initio overlap and Hamiltonian
matrices for a surface silicon atom, $\Sigma_1$ is the desired
self-energy coupling it to the bulk, and $g$ is the silicon surface
Green's function.  We start with an appropriate semi-empirical
description of the bulk silicon bandstructure, and use a real-space,
recursive formalism along the (100) direction to get the semi-empirical
surface Green's function \cite{rManoj}. We then use Eq.~\ref{e00}
to adjust the self-energy $\Sigma_1$ to achieve the best match between
the surface Green's functions (ab-initio and semi-empirical) in
real-space (Fig.~\ref{f0b}), including if required a suitable
reconstructed cluster for unpassivated surfaces. We thus have an
ab-initio description of the molecule and its bonding with the silicon
surface that is consistent with a semi-empirical description of the
bulk and surface properties of the silicon crystal \cite{rLAR}.

We use a recently developed method \cite{rDamle,rDamle2} to calculate
the current, coupling the nonequilibrium Green's function (NEGF)
formulation of transport \cite{rDattabook} with an LDA molecular
Hamiltonian calculated using GAUSSIAN98 \cite{rGauss} in a 6-31g$^*$
basis:
\begin{eqnarray}
G&=&\left[ES - H - \Sigma_1 - \Sigma_2\right]^{-1}\nonumber\\
T&=&{\rm{Trace}}\left[\Gamma_1G\Gamma_2G^\dagger\right],~~\Gamma_{1,2}=i\left[\Sigma_{1,2}
- \Sigma_{1,2}^\dagger\right]\nonumber\\
I &=& (2e/h)\int dE~T(E)\left[f_1(E) - f_2(E)\right]
\end{eqnarray}
where $S,H$ are the molecular overlap and Hamiltonian matrices and
$f_{1,2}$ are the contact Fermi energies.  The calculated current
includes contributions both from thermionic emission over the vacuum
barrier, as well as quantum mechanical tunneling through it.  We
include the effects of an incoherent background inside the molecule
through an
additional phase-randomizing voltage-probe \cite{rButtiker},
incorporated in the generalized Landauer-B\"uttiker formula
\cite{rDattaSL}.

It is well known \cite{rHoffman} that unpassivated silicon has surface
states in the band-gap, as we indeed observe (Fig.~\ref{f0b}).  Some of
these states are eliminated once the molecule bonds to the surface
silicon dimers \cite{rBond}. The surface states in unpassivated silicon
would influence the STM current only if they can be
replenished from the bulk through inelastic processes occuring in the
contact. We ignore such effects in this paper.

In any STM measurement on a molecule, it is important to deconvolve the
tip and the molecular density of states (DOS) so that one can
unambiguously identify features intrinsic to the molecule.  Modeling
tip-sample interactions requires a precise knowledge of the
experimental tip structure, as well as its exact position over the
molecule. It is generally believed \cite{rChen} that the tunneling
current drawn by a tungsten tip is dominated by the d-electrons of
loose tungsten atoms at its end. We model the tip-sample interaction by
evaluating the {\it{largest}} coupling matrix element between the
styrene HOMO wavefunction and the d-electrons of a single tungsten atom
sitting at varying distances from it.  The coupling is obtained using
LDA in a LANL2DZ basis, and yields a broadening $\sim$ 300 meV when the
STM almost touches the molecule and 12 meV when the tip sits about 2
\AA~away. The STM self-energy $\Sigma_2$ is obtained using Fermi's
Golden rule, with a tip DOS $\sim 0.25/$eV. We find that the coupling
can vary considerably depending on the lateral position of the tip atom
relative to the molecule, due to accidental symmetries in the overlap
matrix elements.  This needs further consideration both from theory and
experiment, given that the current level is influenced substantially by
the STM tip.

Much of molecular RTD action hinges on the bias-dependence of the
molecular levels and broadenings. Using the semiconductor bulk band-edges as
reference, the electronic energy levels and wavefunctions, the silicon
band-bending, as well as the tip electrochemical potential vary under
bias (Fig.~\ref{f0}).  Under bias, the molecular levels move relative
to the silicon band-edge, because (1) a large part of the applied bias
drops across the tip-to-substrate gap due to the large dielectric
constant of silicon relative to the molecule and the vacuum gap; and
(2) the field inside the molecule is poorly screened by the absence of
metal-induced gap states (MIGS) that would normally exist with gold
contacts (this creates a low quantum capacitance which causes a
sizeable potential drop within the molecule \cite{rparadigms,rLiang}).
At a critical voltage determined by the electrostatics, the HOMO level
in Fig.~\ref{f0} reaches the bulk silicon valence band-edge (VB) after
which it enters the Si band-gap and its transmission drops
abruptly, leading to a prominent NDR in the I-V of the molecule. Thus
the self-consistent potential $U_{\rm{scf}}$ is
the most important part of the modeling.

The equilibrium band-bending in silicon is given essentially by the
workfunction difference between silicon and tungsten, and is distributed
over a depletion width $\sim$ 1 nm. We obtain the voltage-dependence
of the silicon band-bending using MEDICI \cite{rmedici}, which
agrees with simple estimates based on the depletion-width approximation
\cite{rDW}, as well as surface photovoltage (SPV) measurements
\cite{rHamers}. The band-bending affects the silicon surface Green's
function through the onsite energies of the silicon atoms distributed
over a depletion width. Band-bending turns out however, to be relatively
unimportant, since it is transparent to electron tunneling, so that the
NDR is determined by the {\it{bulk}} silicon band-edges.  The Laplace
part of the potential drop is obtained using MEDICI \cite{rmedici} by
simulating a silicon surface separated from tungsten by a vacuum gap
and a molecular dielectric with a dielectric constant $\sim$ 2. Stark
effects in the molecular DOS are included in GAUSSIAN98 by applying a
molecular field extracted from the Laplace potential. 

\begin{figure}[ht]
\vskip 4.2in
{\includegraphics{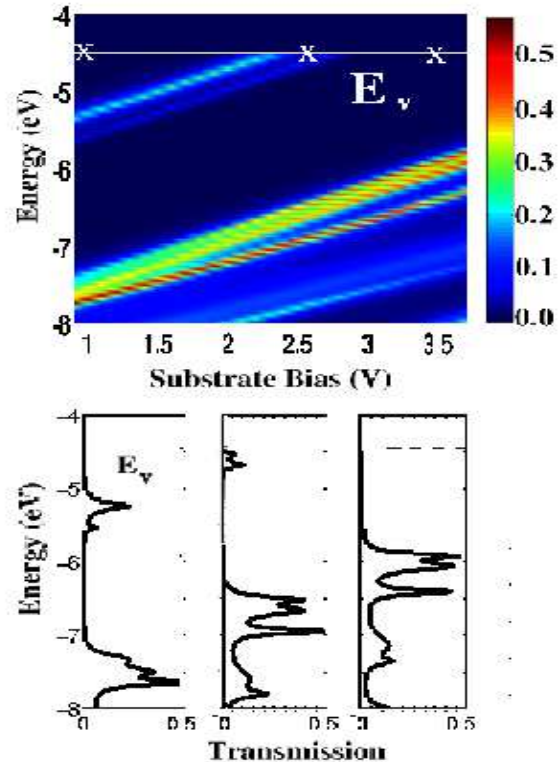}}
\caption{Calculated transmission (colorplot) through styrene 
absorbed on p-silicon. The silicon is hydrogen-passivated and doped at 
$N_A = 5\times 10^{19}$ cm$^{-3}$, giving a Fermi energy 
$E_F \approx 0.1$ eV below its valence bandedge ($E_V$). Styrene HOMO 
peaks shift under bias, and their transmission disappears when they
reach the band-edge $E_V$. The transmission
is plotted for three candidate voltage values, marked by crosses in
the colorplot.}
\label{f3}
\end{figure}

\begin{figure}
\vskip 3.0in
{\includegraphics{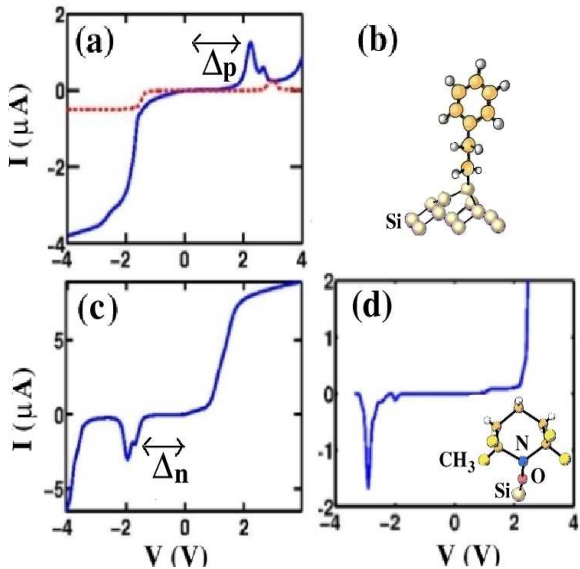}}
\caption{ (a) Calculated
I-V characteristics for styrene on p-doped H-passivated Si(100) with an STM almost touching the molecule.
Increasing the air gap to 2 \AA~(dashed line) changes the overall current and 
postpones the NDR action; (b) the bonding geometry \cite{rWolkow2}
between an end carbon atom on styrene and a c(4x2) Si(100) buckled dimer;
I-Vs  for (c) styrene on n-doped H-passivated Si(100)
and (d) TEMPO on n-doped clean Si(100), $N_D = 5 \times 10^{19}$ cm$^{-3}$.}
\label{f4}
\end{figure}

{\it{Results.}} Eq.~\ref{e00} can be used to obtain the coupling of a
molecule to silicon given a bonding geometry and a surface structure. The
actual bonding geometries vary \cite{rMajor}, and are sometimes
controversial \cite{rbondgeom}, so we will adopt specific models in
this paper for illustrative purposes.  We start by simulating transport
through styrene grown on a p-doped ($N_A = 5 \times 10^{19}$ cm$^{-3}$)
H-passivated Si(100) surface (bonding shown in Fig. ~\ref{f4}b).  Our LDA
energy levels and wavefunctions of styrene agree well with previously
published first-principles results for the above bonding geometry
\cite{rWolkow2}. We use our hybrid basis scheme to connect
this geometry to bulk silicon, described here within an effective mass
model. The effective mass model is used for simplicity and describes
the band-edge properties correctly (Fig.~\ref{f0b}) where most of our
physics lies; future publications \cite{rLAR} will generalize this to more
sophisticated models for bulk silicon.  Fig.~\ref{f3} shows a colorplot
of the transmission of styrene on p-Si under bias.  For increasing
positive substrate voltage the molecular levels increase in energy
until a HOMO level reaches the silicon valence band-edge and its
transmission gets cut-off thereafter. This leads to an NDR in the I-V
(Fig.~\ref{f4}a), calculated for an STM tip almost touching the
molecule. The reduced transmission also leads to an overall current
level that is much smaller for positive substrate bias. Multiple HOMO
levels contribute to multiple NDR peaks.  For negative bias, there is
no such molecular NDR, since the HOMO levels move deeper into the
valence band, and the LUMO levels lie outside the bias window. An
increased tip-to-sample vacuum gap (2 \AA) causes a smaller voltage
drop across the molecule, postponing the onset of the NDR peaks (dashed
line).  The NDR peak position from a level $E_0$ is roughly given by
$(E_V-E_0)/\eta$, where $\eta$ denotes the average molecular potential
under unit applied bias ($\eta \sim 0.5$ for an STM touching the
molecule, decreasing approximately linearly with increasing vacuum
gap).

For molecules on n-silicon, NDR is expected at negative substrate
voltages, mediated through the LUMO levels. The LDA LUMO levels for
styrene and TEMPO are a few volts higher than $E_C$, pushing the
first NDR peak outside the usual bias range.
Fig.~\ref{f4}c shows the results for n-styrene, with the LUMO levels
lowered artificially by about 1.5 eV.  Electron addition levels are
usually controversial, and could differ from theoretical predictions
due to various effects associated with charging, correlation or image
effects. The theoretical NDRs compare well with room temperature
STM measurements
by Hersam {\it{et al.}} \cite{rHersamexpt} on TEMPO and styrene, both
single molecules and monolayers.  The experiments are performed on
clean Si(100) surfaces, with a different surface reconstruction (2x1),
bonding mechanism (cycloaddition), and bandbending \cite{rbandbb},
but the corresponding theoretical results are qualitatively similar
(Fig.~\ref{f4}d).

The origin of NDR in our simulations is qualitatively different from the
NDR mediated by tunneling through localized atomic states arising from
a sharp structure in the tip DOS \cite{rAvouris}.  NDR due to a tip to
level resonance would show up either for both bias directions for both
p and n-type samples (if the tip structure interacts with both HOMO and
LUMO levels), or for one bias direction for only one of the samples (if
it interacts with just one of them). {\it{The consistent observation of a
one-sided NDR, at positive substrate bias in p-Si and negative in n-Si
\cite{rHersamexpt},
can only be explained by interactions between the molecular levels and
the silicon band-edge}}.

The multiple NDR peaks arise from the multiplicity of levels sequentially
crossing the relevant band-edge.  This is a significant improvement
over gold-based experiments, where large molecular charging
energies ($\sim$ 2 eV) prevent an STM electrochemical potential
from crossing more than one level. So far, experimental molecular
conductances exhibit very few (typically zero \cite{rLindsay}, one
\cite{rWeber,rReif} or two \cite{rReed}) peaks, in contrast to systems
with smaller charging energies \cite{rBanin}. In molecular RTDs, however,
the proximity of the band-edge to the Fermi energy dramatically reduces
this charging effect, because any charging due to
electron/hole addition to a level by the stronger coupled substrate is
counterbalanced almost instantly by its removal due to the dramatic
reduction in substrate coupling at the bandedge. The low charging
allows the observation of multiple NDR peaks that can be correlated
with the molecular levels. Indeed, we can use a combination of p and
n-type experiments on a single molecule to map out its HOMO-LUMO
spectrum. For instance, the experimental peak voltages $\Delta_{n,p}$
and Fermi energies $E_{fn,fp}$ for $n$ and $p$-doped substrates give a
HOMO-LUMO gap $\approx \eta\left(\Delta_n + \Delta_p\right) +
(1-\eta)\left(E_{fn} - E_{fp}\right)$, including equilibrium
tip-substrate band-alignment effects.  Charging can also lead to a
hysteresis in the I-V, associated with the changeover of the substrate
coupling from strong to weak at the NDR position. Such hysteresis
effects are observed in RTDs \cite{rRTDhyst}, but will also be
compromised by the low charging in these molecules. For realistic
device parameters, the predicted hysteresis is small, and could be
washed out by scattering.

We are grateful to M. Hersam and his group for liberally sharing their
unpublished experimental results with us. In addition, we would like to
thank M. S. Lundstrom, C-F. Huang,  D. Janes and M. Paulsson for useful
discussions.  This work has been supported by ARO-DURINT, Award\#
527826-02, NSF, Award\# 0085516-EEC and by the Semiconductor Technology
Focus Center on Materials, Structures and Devices under contract \#
1720012625.


\begin{thebibliography}{100}
\bibitem{rLAR} Detailed results for bulk and surface structures on silicon
will be described in a forthcoming publication.
\bibitem{rbandbb} The band-bending is constrained to lie between $\pm 0.2$
V by localized $\pi$ and $\pi^*$ dimer levels, as confirmed by SPV
measurements (\cite{rHamers}). This makes it easier for the molecular levels to 
``slip'' past the silicon surface bandedge, although it is the bulk bandedge that
principally determines the NDR positions.
\end{thebibliography}
\end{document}